\documentclass[aip,amsmath,amssymb,reprint]{revtex4-1}

\usepackage{enumitem}

\usepackage{spverbatim}
\usepackage{graphicx}% Include figure files
\usepackage{dcolumn}% Align table columns on decimal point
\usepackage{bm}% bold math
\usepackage[utf8]{inputenc}
\usepackage[T1]{fontenc}
\usepackage{newtxtext}
\usepackage{newtxmath}
\usepackage{etoolbox}

\usepackage{fancyvrb}    % for spverbatim
\usepackage{mdframed}    % for colored framed boxes
\usepackage{xcolor}      % for defining colors

\definecolor{lightgray}{gray}{0.95} % adjust 0.9-0.97 for lighter/darker

\newmdenv[
  backgroundcolor=lightgray,
  linecolor=lightgray,     % no visible border (same color as background)
  skipabove=10pt,
  skipbelow=10pt,
  innerleftmargin=10pt,
  innerrightmargin=10pt,
  innertopmargin=8pt,
  innerbottommargin=8pt
]{grayverb}
\makeatletter
\def\@email#1#2{%
 \endgroup
 \patchcmd{\titleblock@produce}
  {\frontmatter@RRAPformat}
  {\frontmatter@RRAPformat{\produce@RRAP{*#1\href{mailto:#2}{#2}}}\frontmatter@RRAPformat}
  {}{}
}%
\makeatother
\begin{document}

% \preprint{AIP/123-QED}

\title{fix pimd/langevin: An Efficient Implementation of Path Integral Molecular Dynamics in LAMMPS}

\author{Yifan Li}\thanks{Email: yifanl0716@gmail.com}
\affiliation{ 
Department of Chemistry, Princeton University, Princeton, NJ 08544, USA
}
\author{Axel Gomez}
\affiliation{ 
Department of Chemistry, Princeton University, Princeton, NJ 08544, USA
}
\author{Kehan Cai}
\affiliation{ 
Department of Chemistry, Princeton University, Princeton, NJ 08544, USA
}
\author{Chunyi Zhang}
\affiliation{ 
Eastern Institute of Technology, Ningbo, Zhejiang 315200, China
}
\author{Li Fu}
\affiliation{ 
School of Materials Science and Engineering, Peking University, Beijing 100871, China}
\author{Weile Jia}
\affiliation{ 
University of Chinese Academy of
Sciences, Beijing 101408, China}
\author{Yotam M. Y. Feldman}
\affiliation{School of Chemistry, Tel Aviv University, Tel Aviv 6997801, Israel}
\author{Ofir Blumer}
\affiliation{School of Chemistry, Tel Aviv University, Tel Aviv 6997801, Israel}
\author{Jacob Higer}
\affiliation{School of Physics, Tel Aviv University, Tel Aviv 6997801, Israel}
\author{Barak Hirshberg}
\affiliation{School of Chemistry, Tel Aviv University, Tel Aviv 6997801, Israel}
\affiliation{The Center for Computational Molecular and Materials Science, Tel Aviv University, Tel Aviv 6997801, Israel}
\author{Shenzhen Xu}
\affiliation{
School of Materials Science and Engineering, Peking University, Beijing 100871, China}
\author{Axel Kohlmeyer}
% \ead{axel.kohlmeyer@temple.edu}
\affiliation{Institute for Computational Molecular Science, Temple University, Science Education and Research Center (035-07), Philadelphia, PA 19122, USA}
\author{Roberto Car}
\affiliation{ 
Department of Chemistry, Princeton University, Princeton, NJ 08544, USA
}
\affiliation{ 
Program in Applied and Computational Mathematics, Princeton University, Princeton, NJ 08544, USA%\\This line break forced with \textbackslash\textbackslash
}
\affiliation{ 
Department of Physics, Princeton University, Princeton, NJ 08544, USA%\\This line break forced with \textbackslash\textbackslash
}%
\affiliation{ 
Princeton Institute for the Science and Technology of Materials, Princeton University, Princeton, NJ 08544, USA%\\This line break forced with \textbackslash\textbackslash
}%

\date{\today}% It is always \today, today,

\begin{abstract}
Path integral molecular dynamics (PIMD), which maps a quantum particle onto a fictitious classical system of ring polymers and propagates the ``beads'' of this extended classical system using molecular dynamics, is widely used to capture nuclear quantum effects (NQEs) in molecular simulations. Accurate PIMD calculations typically require a large number of beads and are therefore computationally demanding. While software packages such as i-PI offer comprehensive PIMD functionality, the high efficiency of simulations driven by machine learning interatomic potentials, such as Deep Potential (DP), calls for more efficient PIMD implementations that fully exploit modern massively parallel supercomputers. Here we present \texttt{fix pimd/langevin}, an efficient PIMD implementation in LAMMPS that supports commonly used features and leverages the Message Passing Interface architecture of LAMMPS to achieve high computational efficiency. We demonstrate the usage and validate the correctness of our code using liquid water as a representative example, and provide a comprehensive overview of the supported features. Then we discuss several important technical aspects of the implementation. Using DP simulations of water as a benchmark, we show that our implementation achieves several-fold acceleration compared to i-PI. Finally, we report strong and weak scaling results that demonstrate the favorable parallel performance of our code.
\end{abstract}

\maketitle

\section{Introduction}
Path integral molecular dynamics (PIMD) is a widely used approach for modeling the quantum mechanical behavior of atomic nuclei~\cite{marx_ab_1996,ceriotti_nuclear_2016,markland_nuclear_2018}. Nuclear quantum effects (NQEs) play an important role in a broad range of phenomena, where isotope effects are crucial for thermodynamic properties like ice melting temperature~\cite{li_assessment_2025,li_ab_2025} and isotope fractionation between mineral and aqueous fluids~\cite{gao_path-integral_2023}, or chemical reactivity such as water dissociation at solid-liquid interfaces~\cite{cao_quantum_2025} and proton transport in water~\cite{gomez_neural-network-based_2024}. PIMD is an invaluable tool for simulating these effects and has therefore been implemented in many widely used molecular dynamics (MD) software packages, including OpenMM~\cite{eastman_openmm_2010,eastman_openmm_2017}, AMBER~\cite{case_amber_2005}, GPUMD~\cite{ying_highly_2025,xu_gpumd_2025}, and i-PI~\cite{ceriotti_i-pi_2014,kapil_i-pi_2019,litman_i-pi_2024}. Performing PIMD simulations with i-PI has become common practice due to its rich feature set and versatile interfaces with a wide range of force engines.

In recent years, the development of machine learning interatomic potentials (MLIPs)~\cite{behler_generalized_2007, zhang_deep_2018, cheng_ab_2019, xu_isotope_2020} has paved the way for long-time PIMD simulations of large systems with \textit{ab initio} accuracy~\cite{gomez_neural-network-based_2024}. Simulations of bosonic systems have also become accessible through efficient polynomial-time PIMD algorithms~\cite{hirshberg_path_2019,feldman_quadratic_2023,higer_periodic_2025}. However, existing PIMD implementations can become performance bottlenecks in large-scale simulations, particularly when MLIP force evaluations are sufficiently fast. For example, although substantial effort has been devoted to improving the performance of i-PI in its v3.0 release~\cite{litman_i-pi_2024}, its Python-based serial architecture and client-server communication scheme still introduce intrinsic overhead, limiting its applicability to large systems. Moreover, the existing PIMD implementation provided by LAMMPS~\cite{plimpton_fast_1995,thompson_lammps_2022} is no longer actively maintained and lacks several essential features, such as support for the $NpT$ ensemble, which restricts its practical applicability. In contrast, LAMMPS is designed for large-scale parallel simulations and supports a wide range of force fields and MLIPs, making it one of the most widely used MD software packages. Therefore, a comprehensive and optimized PIMD implementation directly within LAMMPS is an attractive target for development.

Here, we present an efficient PIMD implementation in LAMMPS, named ``\texttt{fix pimd/langevin} '', which has been included in the official LAMMPS distribution since the feature release of June 15, 2023. This implementation incorporates commonly used PIMD functionalities and is optimized for high performance through MPI parallelization. The two-level parallelization strategy of LAMMPS PIMD allows each bead to be parallelized independently while enabling GPU or OpenMP acceleration for each bead, provided that the underlying force field supports it. When combined with highly efficient machine-learning interatomic potentials (MLIPs), it enables large-scale PIMD simulations with \textit{ab initio} accuracy on massively parallel supercomputers.

The remainder of this paper is organized as follows. In Section II, we briefly review the theoretical foundations of PIMD. In Section III, we describe the usage of the code by presenting an example of liquid-water simulations, discussing the treatment of long-range electrostatic interactions in PIMD, and summarizing the supported features. Section IV details several technical aspects of the implementation. In Section V, we benchmark the performance of our implementation when interfaced with the Deep Potential (DP)~\cite{zhang_deep_2018,wang_deepmd-kit_2018,zeng_deepmd-kit_2023} MLIP, including a comparison with i-PI and an analysis of the strong and weak scaling behavior. Finally, Section VI provides concluding remarks and perspectives.

\section{Theory}
PIMD is based on Feynman's imaginary-time path integral formulation of quantum statistical mechanics. In PIMD, a quantum system of $N$ atoms is mapped onto a ring polymer system consisting of $n$ beads of these $N$ atoms and classical MD is used to evolve this ring polymer configuration with the Hamiltonian

\begin{widetext}
\begin{equation}\label{Hamiltonian}
H=\sum_{k=0}^{n-1}\left(\sum_{i=1}^{N}\frac{1}{2}m_i\bm{v}_{i}^{(k)2}+\sum_{i=1}^{N}\frac{1}{2}m_i\omega_n(\bm{r}_{i}^{(k)}-\bm{r}_{i}^{(k+1)})^2+U(\bm{r}_{1}^{(k)},\dots,\bm{r}_{N}^{(k)})\right)
\end{equation}
\end{widetext}
with the condition $\bm{r}_{i}^{(0)}=\bm{r}_{i}^{(n)}$. Here $\bm{r}_{i}^{(k)}$ and $\bm{v}_{i}^{(k)}$ denote the coordinates and velocities of the $i$-th atom of the $k$-th bead, respectively, and $m_i$ is the mass of the $i$-th atom. The frequency of the harmonic springs in the ring polymer system is
\begin{equation}\label{omega_n}
\omega_n = \frac{n}{\beta \hbar},\quad\beta = 1/(k_{\mathrm{B}}T),
\end{equation}
where $k_{\mathrm{B}}$ is Boltzmann's constant and $\hbar$ is the reduced Planck's constant. $U(\bm{r}_{1}^{(k)},\dots,\bm{r}_{N}^{(k)})$ denotes the potential energy of the $k$-th bead.

The static equilibrium properties of the quantum system can then be computed from appropriate estimators evaluated over the trajectory of the beads. We list the expressions of some important estimators we have implemented in Appendix~\ref{app:properties}. For a pedagogical introduction to PIMD, we refer the readers to Chapter 12 of Ref.~\citenum{tuckerman_statistical_2023}.

The ring polymer formulation introduces an additional set of high-frequency normal modes. To efficiently thermostat these normal modes, Tuckermann et al.~\cite{tuckerman_efficient_1993} established the deterministic Nos\'e-Hoover chain (NHC) thermostat for PIMD. Ceriotti et al.~\cite{ceriotti_efficient_2010} developed the stochastic path integral Langevin equation (PILE) thermostat, which has the local (PILE\_L) and global (PILE\_G) variants. In this implementation we focus on the PILE\_L thermostat and thus call our implementation \texttt{fix pimd/langevin}.

To enable PIMD in the $NpT$ ensemble, Martyna et al.~\cite{martyna_molecular_1999} extended the Martyna-Tuckerman-Tobias-Klein (MTTK) barostat to the context of PIMD. Ceriotti et al.~\cite{ceriotti_i-pi_2014} combined ideas from the PILE thermostat, the stochastic barostat~\cite{bussi_isothermal-isobaric_2009}, and the deterministic barostat for PIMD~\cite{martyna_molecular_1999} to derive a barostat compatible with the stochastic thermostat. Following the convention adopted in i-PI, we refer to this scheme as the Bussi-Zykova-Parrinello (BZP) barostat. In our implementation, we support both the MTTK and BZP barostats.

Several advanced PIMD methods to reduce the number of beads have been developed, such as ring polymer contraction~\cite{markland_efficient_2008} and the colored-noise thermostat~\cite{ceriotti_accelerating_2011}. 
Our implementation can be easily extended to support these methods in the future. Moreover, while quantum dynamics approaches such as centroid molecular dynamics (CMD)~\cite{cao_formulation_1994-2, cao_formulation_1994-3, cao_formulation_1994, cao_formulation_1994-4, cao_formulation_1994-1} and ring polymer molecular dynamics (RPMD)~\cite{craig_quantum_2004} are not discussed here, these methods can be realized within the present framework by selecting appropriate ensembles and thermostat parameters.

\section{Software Usage}
We introduce the usage of the \texttt{fix pimd/langevin} module by presenting an example PIMD simulation of liquid water. We then discuss the incorporation of long-range electrostatic interactions into PIMD. Finally, we provide a list of the features supported by this implementation.
\subsection{Example: PIMD Simulation of Liquid Water}\label{example_liquid}
Here we present PIMD simulations of liquid water as an example usage of \texttt{fix pimd/langevin}. We use a compressed DP model~\cite{lu_dp_2022} trained in Ref.~\citenum{li_assessment_2025} with DFT energies and forces based on the SCAN\cite{sun_strongly_2015} functional to drive the PIMD simulations of a box of 128 H$_2$O molecules using 32 beads in the $NpT$ ensemble. The temperature is kept at 300 K using the PILE\_L thermostat with a damping time 0.1 ps. The pressure is kept at 1 bar using the BZP thermostat with a damping time 0.5 ps. The develop branch of LAMMPS (commit b1ef9f, dated February 11, 2026) is used. The LAMMPS PIMD simulation spans 5 ns with a timestep 0.5 fs. For comparison, additional PIMD simulations are performed using i-PI, interfaced with DeePMD-kit via the \texttt{dp\_ipi} interface, employing the same parameters. We run five independent i-PI simulations with different initial configurations, each spanning 1 ns. The PIMD simulations achieve speeds of 7.0 ns/day with LAMMPS and 1.4 ns/day with i-PI, respectively. The input files and the DP model are available in the companion GitHub repository~\cite{li_httpsgithubcomyi-fanlilammps_fix_pimd_langevin_nodate}. 

Running PIMD in LAMMPS relies on its ``partition'' feature based on a two-level MPI parallelization~\cite{thompson_lammps_2022}. A PIMD task for $n$ beads runs on $N=n\times M$ processors, where the $N$ processors are divided into $n$ partitions with $M$ processors. Each bead runs on $M$ processors and the information of different beads is exchanged through inter-bead MPI communications. The user needs to allocate $N$ processors and run the job with the command \texttt{mpirun -np N lmp -in in.lmp -p n$\times$M}. Notably, the number of beads $n$ is set in this command rather than in the input file. In the input script, one needs to specify the PIMD parameters by defining a \texttt{fix pimd/langevin} line. The following line is an example: 

\begin{grayverb}
\begin{spverbatim}
fix 1 all pimd/langevin method nmpimd integrator obabo ensemble npt temp 300.0 thermostat PILE_L 1234 tau 1.0 scale 1.0 iso 1.0 barostat BZP taup 1.0 fixcom no
\end{spverbatim}
\end{grayverb}
\vspace{1em}  

This command defines a PIMD simulation using the normal mode representation and the OBABO Trotter splitting scheme~\cite{tuckerman_statistical_2023,ceriotti_efficient_2010}. The keyword \texttt{ensemble npt} specifies an $NpT$ simulation. The option \texttt{temp 300.0} sets the temperature to 300~K, which is controlled using the PILE\_L thermostat with a random number generator seed of \texttt{1234}. The parameter \texttt{tau 1.0} sets the damping time of the centroid mode to 1.0~ps, while \texttt{scale 1.0} assigns the damping times of the non-centroid modes to their optimal values, as defined in Eq.~(36) of Ref.~\citenum{ceriotti_efficient_2010}. The option \texttt{iso 1.0} applies an isotropic pressure of 1~bar, controlled by the BZP barostat with a damping time of 1.0~ps specified by \texttt{taup 1.0}. Finally, \texttt{fixcom no} indicates that center-of-mass velocities are not removed during the simulation.

\texttt{fix pimd/langevin} outputs several useful thermodynamic properties, as listed in Appendix~\ref{app:properties} and the documentation. For example, the seventh output corresponds to $K_{\mathrm{CV}}$, the centroid-virial kinetic energy estimator, as defined in Eq.~\eqref{eq_kcv}. The tenth output corresponds to $P_{\mathrm{CV}}$, the centroid-virial pressure estimator, as defined in Eq.~\eqref{eq_pcv}. One can print $K_{\mathrm{CV}}$ and $P_{\mathrm{CV}}$ from PIMD simulations by adding the following lines to the input file:

\begin{grayverb}
\begin{spverbatim}
thermo_style custom step f_1[7] f_1[10]
thermo_modify format step %ld format 2* "%.16e"
\end{spverbatim}
\end{grayverb}
\vspace{1em}  

The trajectory of each bead is output to a separate file. To label each trajectory file, we first define a variable \texttt{ibead}, which represents the bead ID, at the beginning of the input file:

\begin{grayverb}
\begin{spverbatim}
variable ibead uloop 99 pad
\end{spverbatim}
\end{grayverb}
Thus, the atomic coordinates can be dumped to files named \texttt{\$\{ibead\}.xyz}:
\begin{grayverb}
\begin{spverbatim}
dump 1 all custom 100 \${ibead}.xyz id type x y z
dump_modify 1 sort id
\end{spverbatim}
\end{grayverb}
\vspace{1em}
We point out that the velocities (\texttt{vx}, \texttt{vy}, \texttt{vz}) and forces (\texttt{fx}, \texttt{fy}, \texttt{fz}) output by \texttt{fix pimd/langevin} are given in the normal mode representation, whereas the coordinates (\texttt{x}, \texttt{y}, \texttt{z}) are output in Cartesian coordinates.

\begin{figure}[ht!]
     \centering
\includegraphics[width=\columnwidth]{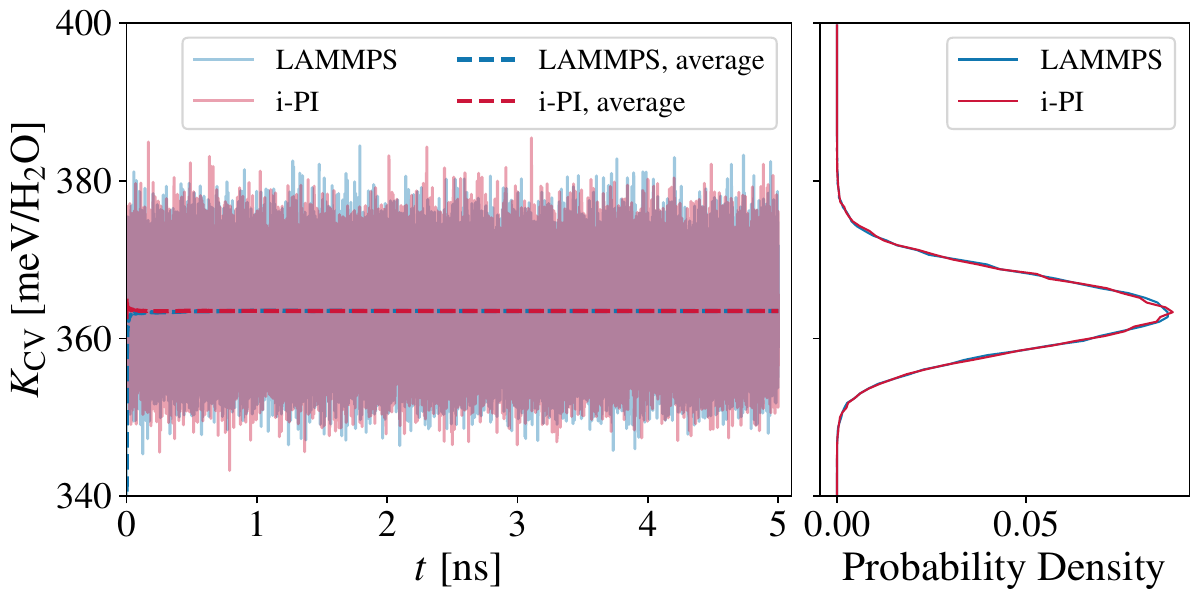}
\caption{Centroid-virial quantum kinetic energy estimator $K_{\mathrm{CV}}$. Left: trajectory of the quantity (solid line) and its cumulative average (dashed line). Right: corresponding distribution. Blue lines correspond to LAMMPS results and red lines to i-PI results.}
\label{kcv}
\end{figure}

\begin{figure}[ht!]
     \centering
\includegraphics[width=\columnwidth]{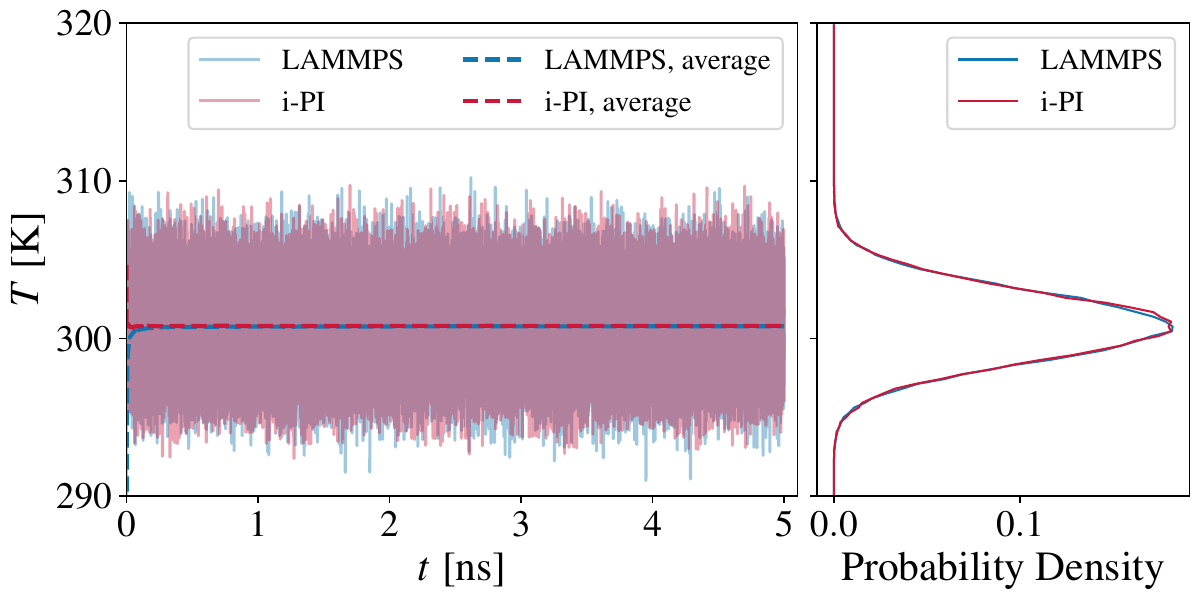}
\caption{Temperature $T$ of the classical ring polymer system. The temperature printed by LAMMPS's \texttt{temp} should be divided by the number of beads for a proper comparison. Left: trajectory of the quantity (solid line) and its cumulative average (dashed line). Right: corresponding distribution. Blue lines correspond to LAMMPS results and red lines to i-PI results.}
\label{temperature}
\end{figure}

\begin{figure}[ht!]
     \centering
\includegraphics[width=\columnwidth]{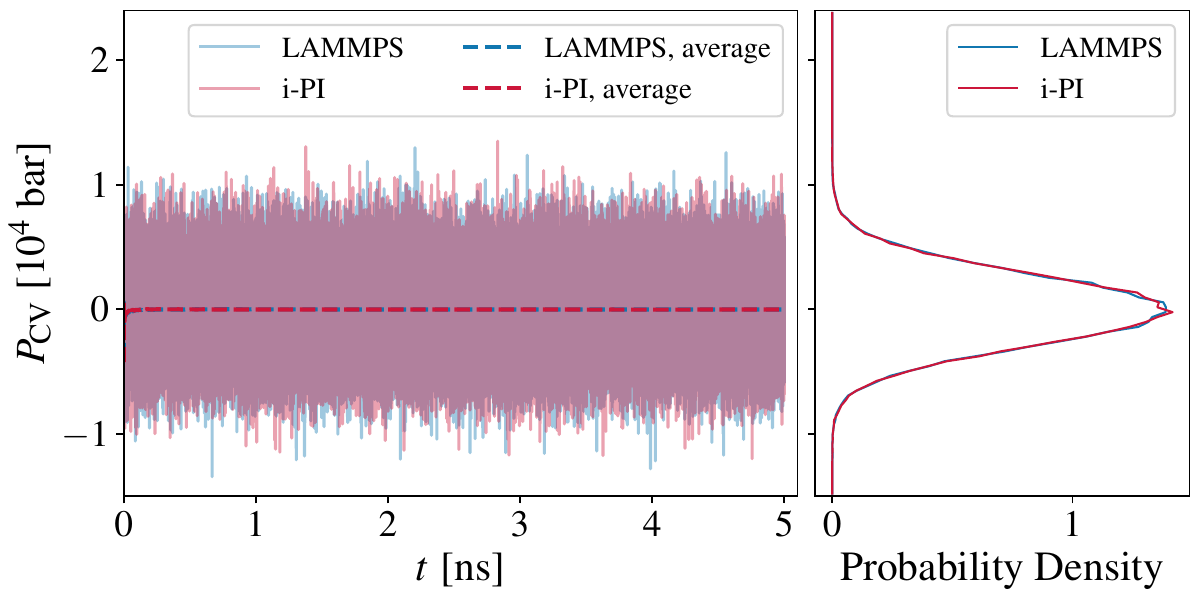}
\caption{Centroid-virial pressure estimator $P_{\mathrm{CV}}$. Left: trajectory of the quantity (solid line) and its cumulative average (dashed line). Right: corresponding distribution. Blue lines correspond to LAMMPS results and red lines to i-PI results.}
\label{pcv}
\end{figure}

\begin{figure}[ht!]
     \centering
\includegraphics[width=\columnwidth]{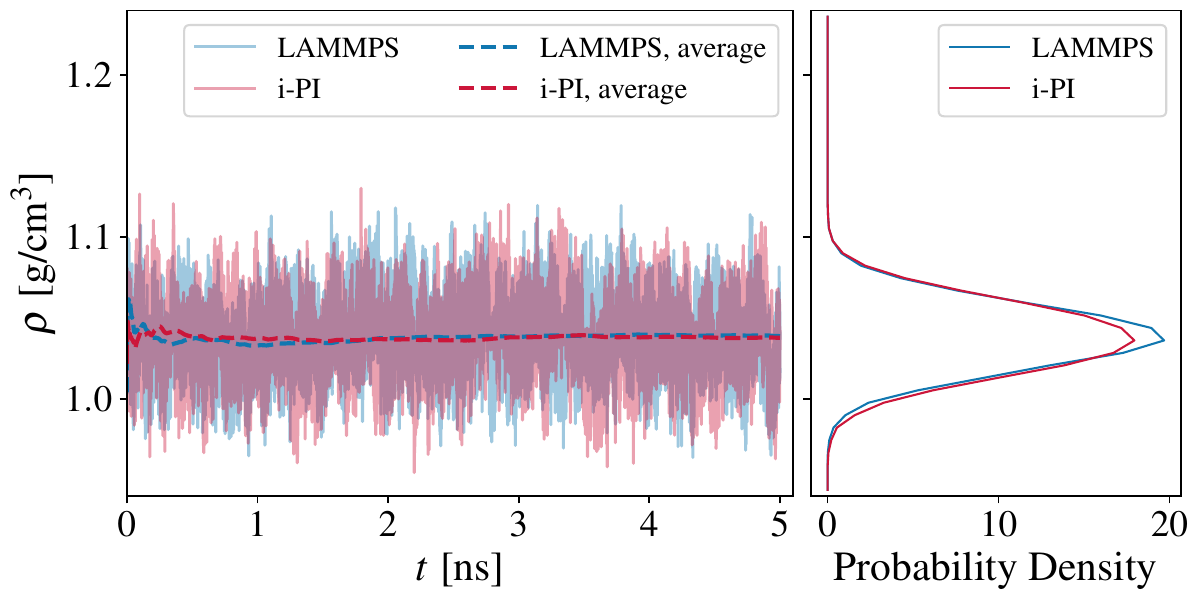}
\caption{The density $\rho$ of the system. Left: trajectory of the quantity (solid line) and its cumulative average (dashed line). Right: corresponding distribution. Blue lines correspond to LAMMPS results and red lines to i-PI results.}
\label{density}
\end{figure}

We present the trajectories and distributions of $K_{\mathrm{CV}}$, the ring polymer system temperature $T$, $P_{\mathrm{CV}}$, and the density $\rho$ from our PIMD simulations in FIGs.~\ref{kcv}, \ref{temperature}, \ref{pcv}, and \ref{density}, respectively. These results show that our LAMMPS implementation samples the correct distributions for these thermodynamic quantities, in full agreement with i-PI, thereby validating the correctness of our implementation.

\begin{figure}
     \centering
\includegraphics[width=\columnwidth]{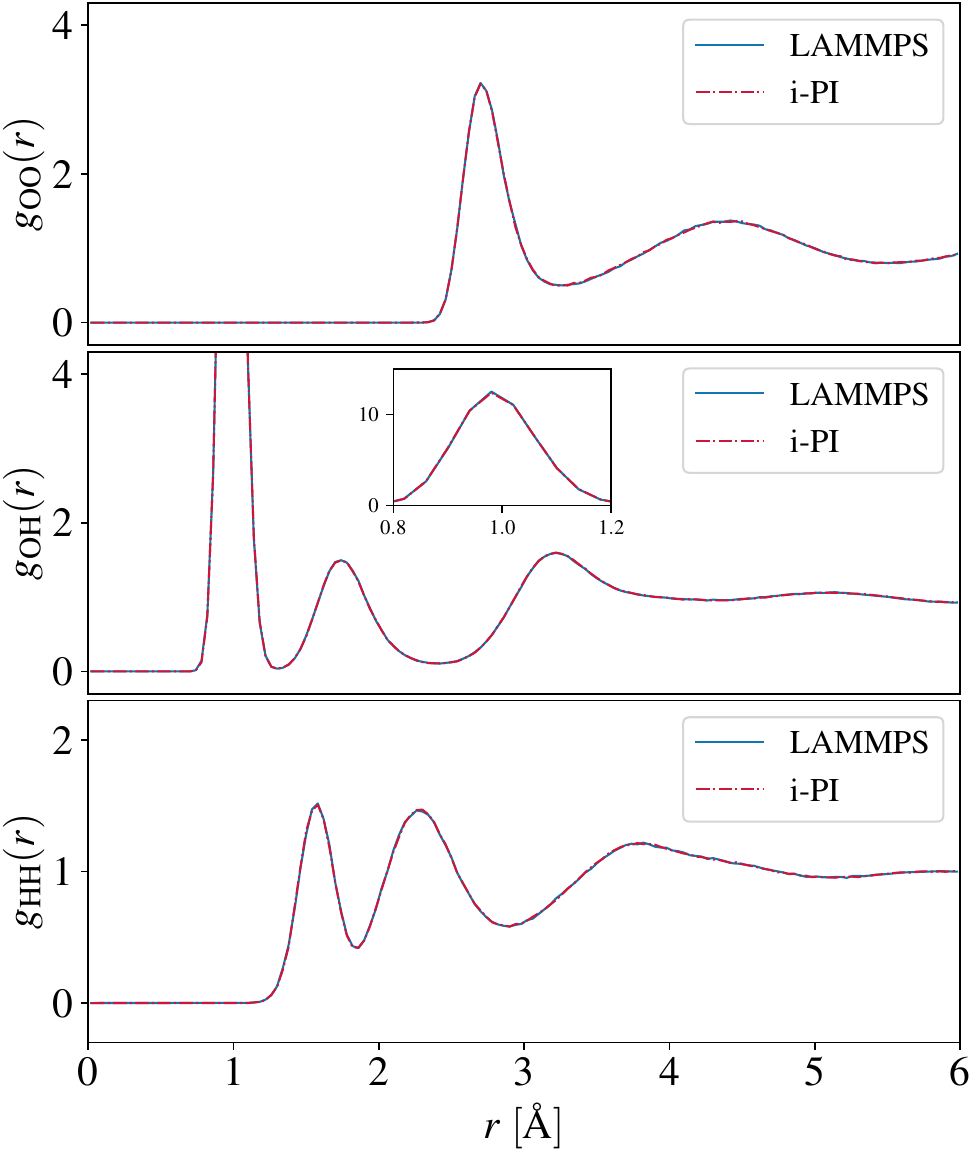}
\caption{The radial distribution functions (RDFs) of liquid water from PIMD simulations for 128 H$_2$O molecules at 300 K and 1 bar. The blue line corresponds to LAMMPS results and the red line to i-PI results.}
\label{RDF-water}
\end{figure}

Moreover, we show the O-O, O-H, and H-H radial distribution functions (RDFs) $g_{\mathrm{OO}}(r)$, $g_{\mathrm{OH}}(r)$, and $g_{\mathrm{HH}}(r)$ of liquid water in Figure \ref{RDF-water}. The RDFs from LAMMPS and i-PI are consistent, further demonstrating the correctness of our LAMMPS implementation. 

\subsection{PIMD with DP Long-Range}
We now describe how to run a PIMD simulation driven by the Deep Potential long-range (DPLR) model~\cite{zhang_deep_2022},  which explicitly incorporates long-range electrostatic interactions into the DP framework. Including long-range electrostatics is particularly important for inhomogeneous systems (e.g., clusters, vapor phase, and interfaces) and for properties that are sensitive to electrostatic interactions, such as dielectric response. Neglecting long-range electrostatic interactions has been shown to produce spurious exponential decay of dipolar correlations, inaccurate free-energy landscapes, or artificial violations of charge neutrality~\cite{calegari_andrade_probing_2023,zhang_molecular-scale_2024,zhang_tuning_2025,cai_simulations_2025,zhang_infrared_2025}.
A complete example of a DPLR-driven PIMD simulation of liquid water is provided in our GitHub repository~\cite{li_httpsgithubcomyi-fanlilammps_fix_pimd_langevin_nodate}.

The DPLR method explicitly accounts for long-range Coulomb interactions between positively charged ions and negatively charged Wannier centroids (WCs)~\cite{zhang_deep_2022}, which correspond to the average positions of maximally localized Wannier centers (MLWCs). The WC positions are predicted by a separate Deep Wannier (DW) model~\cite{zhang_deep_2020}. In LAMMPS simulations, each WC is represented as an atom whose coordinates are not updated
by the MD integrator but are instead inferred from the DW model. When computing instantaneous thermodynamic quantities, such as temperature or pressure, which are used by some thermostats (e.g., NHC) or barostats (e.g., BZP) to drive the dynamics, only ionic contributions should be included.

To exclude WCs when necessary, we define two groups of atoms in the LAMMPS input file:
\begin{grayverb}
\begin{spverbatim}
group real_atom type 1 2
group virtual_atom type 3
\end{spverbatim}
\end{grayverb}
\vspace{1em}
where \texttt{real\_atom} and \texttt{virtual\_atom} denote ions and WCs, respectively. Atom types 1, 2, and 3 correspond to O, H, and WC. The \texttt{bond\_style}, \texttt{kspace\_style}, and \texttt{fix dplr} commands should also be specified appropriately, following DPLR-driven MD simulations with classical nuclei. Detailed instructions can be found in the DeePMD-kit documentation.

The \texttt{fix pimd/langevin} command should be applied only to the ionic degrees of freedom:
\begin{grayverb}
\begin{spverbatim}
fix 1 real_atom pimd/langevin method nmpimd integrator obabo ensemble npt
temp 350.0 thermostat PILE_L 1234 tau 1.0
iso 1.0 barostat BZP taup 1.0
\end{spverbatim}
\end{grayverb}
\vspace{1em}
Failure to exclude WCs in this command leads to incorrect instantaneous pressure calculations, erroneous $NpT$ trajectories, and incorrect property outputs, such as $K_{\mathrm{CV}}$, although it does not affect the PILE\_L thermostat.

The 32-bead PIMD simulation of liquid water using the DPLR model, with a timestep of 0.5~fs, achieves 0.16~ns/day for a system of 128 H$_2$O molecules on a single Perlmutter node equipped with four NVIDIA A100 GPUs. In comparison, the same 32-bead PIMD simulation driven by the short-range DP model yields 8.6~ns/day on one node. Using two nodes (eight A100 GPUs in total), the DPLR-driven 32-bead PIMD simulation reaches 0.22 and 0.093~ns/day for systems containing 128 and 1024 H$_2$O molecules, respectively. The reduced speed of DPLR simulations compared to those driven by short-range DP models can be attributed to the additional cost induced by reciprocal space calculations of long-range interactions and the predictions of the WCs.

\subsection{List of Features}
Here we list the features currently supported by our implementation. A more comprehensive explanation of these features can be found in the LAMMPS documentation. The values shown in bold font indicate the default values of the corresponding keywords.

\begin{itemize}
    \item \texttt{method} (the coordinate system to integrate the equation of motion): \texttt{\textbf{nmpimd}} (the normal mode transformation defined in Appendix~\ref{app:normalmode} is used); \texttt{pimd} (the Cartesian coordinate is used).
    
    \item \texttt{integrator} (the scheme of the Trotter splitting): \texttt{\textbf{OBABO}} (``O'', ``B'', and ``A'' represent thermostatting, updating the velocities, and updating the coordinates, respectively); \texttt{BAOAB} (the BAOAB integrator appears to improve the numerical stability~\cite{liu_simple_2016}).
    
    \item \texttt{ensemble}: \texttt{NVE} (the microcanonical ensemble); \texttt{\textbf{NVT}} (the canonical ensemble); \texttt{NPH} (the isoenthalpic-isobaric  ensemble); \texttt{NPT} (the isothermal-isobaric ensemble).
    
    \item \texttt{sp}: the scaling factor of Planck's constant. \texttt{sp} can be used to control the ``quantumness'' of the PIMD simulation. In unit styles other than \texttt{lj}, setting \texttt{sp} to \texttt{1} and \texttt{0} corresponds to the fully quantum and classical limits, respectively. For \texttt{lj} units, a fully quantum simulation translates into setting sp to the de Boer quantumness parameter~\cite{de_boer_chapter_1957} $\Lambda^*$ (see Appendix~\ref{deboer} for more details). Default is \texttt{\textbf{1}}.
    
    \item \texttt{fmmode} (mode for selecting fictitious masses of the normal modes): \texttt{\textbf{physical}} (using physical particle masses $m_i^{(k)}=m_i$); \texttt{normal} (using normal mode masses $m_i^{(k)}=\lambda_k m_i$, where $\lambda_k$ is defined in Eq.~\eqref{lambda_k}).
    
    \item \texttt{fmass} (the scaling factor of the fictitious mass): a number that multiplies the particle mass. Setting \texttt{fmass} to $x$ is equivalent to setting \texttt{sp} to $1/\sqrt{x}$. Default is \texttt{\textbf{1}}.
    
    \item \texttt{temp}: the target temperature of the thermostat. Default is \texttt{\textbf{298.15}}.
    
    \item \texttt{thermostat}: \texttt{PILE\_L} (the local path integral Langevin equation thermostat; requires a random \texttt{seed}). Default is \texttt{\textbf{-1}}.
    % ; \texttt{NHC} (the Nos\'e-Hoover chain thermostat).
    % \item \texttt{tcahin}: the number of chains in the Nos\'e-Hoover chain thermostat.
    % \item \texttt{tloop}: the number of sub-cycles to perform on the Nos\'e-Hoover chain thermostat.
    
    \item \texttt{tau}: the damping time for the centroid mode in the PILE\_L thermostat. Default is \texttt{\textbf{1.0}}.
    
    \item \texttt{scale}: a scaling factor applied to the damping times of the non-centroid modes in the PILE\_L thermostat. By default, the damping times of the non-centroid modes are assigned according to Eq.~(36) of Ref.~\citenum{ceriotti_efficient_2010}, and this parameter uniformly rescales those values (see Appendix~\ref{PILE_L_thermostat} for details). Default is \texttt{\textbf{1.0}}.
    
    \item \texttt{\textbf{iso}} or \texttt{aniso} or \texttt{x} or \texttt{y} or \texttt{z}: followed by a target pressure for the barostat. The pressure can be controlled isotropically (using \texttt{iso}), anisotropically (using \texttt{aniso}), or independently along each Cartesian direction (using \texttt{x}, \texttt{y}, or \texttt{z}). Default is \texttt{\textbf{1.0}}.
    
    \item \texttt{barostat}: \texttt{\textbf{BZP}} (the Bussi-Zykova-Parrinello barostat); \texttt{MTTK} (the Martyna-Tuckerman-Tobias-Klein barostat).
    
    \item \texttt{taup}: the damping time of the barostat. Default is \texttt{\textbf{1.0}}.
    
    \item \texttt{fixcom} (whether to remove center-of-mass motion): \texttt{\textbf{yes}}; \texttt{no}. When set to \texttt{yes}, the center-of-mass velocity is removed in every timestep for the centroid mode (when \texttt{method} is \texttt{nmpimd}) or for each bead individually (when \texttt{method} is \texttt{pimd}).
    
\end{itemize}

PIMD simulations of bosons~\cite{hirshberg_path_2019}, including exchange effects, can be performed with the \texttt{fix pimd/langevin/bosonic} command. Integration is performed in Cartesian coordinates only, because the normal modes change under bosonic exchange, and the supported ensembles are $NVE$ and $NVT$. Bosonic kinetic energy estimators are available as property outputs (see Appendix~\ref{app:properties}).

\section{Implementation Details}
In this Section we discuss several technical aspects of our implementation. 

\subsection{Equations of Motion and Integrators}
Within the LAMMPS framework, we implement PIMD in the normal mode representation using the same equations of motion (EOMs) as those adopted by i-PI~\cite{ceriotti_i-pi_2014,kapil_i-pi_2019,litman_i-pi_2024}. These EOMs employ an orthogonal normal mode transformation, as defined in Appendix~\ref{app:normalmode}. In each timestep, we perform forward and inverse normal mode transformations of the coordinates before and after propagating the atomic positions, and apply a normal mode transformation of the forces following the force evaluation.

In the $NVE$ ensemble, the system evolves with a conserved Hamiltonian $H$ as defined in Eq.~\eqref{tot_energy}. To sample the $NVT$ ensemble, we implement the PILE\_L thermostat using the integrator described in Appendix~\ref{PILE_L_thermostat}. Within this framework, a target temperature of $nT$ is applied to thermostat the ring polymer system, where $n$ is the number of beads.

Evolving the extended system of atoms and simulation cell with a conserved generalized enthalpy $\mathcal{H}$ in Eq.~\eqref{total_enthalpy} yields the $NpH$ (isoenthalpic-isobaric) ensemble. The BZP barostat is implemented using the integrator described in Appendix~\ref{BZP_barostat}. Applying a white-noise Langevin thermostat to the barostat velocity and the PILE\_L thermostat to the atomic velocities generates the $NpT$ ensemble.

\subsection{Periodic Boundary Conditions}
Periodic boundary conditions (PBCs) in PIMD simulations within LAMMPS require careful handling. LAMMPS stores atomic coordinates wrapped inside the simulation box and tracks crossings of periodic boundaries using per-atom \texttt{image} flags, which record how many box lengths must be added or subtracted to recover the unwrapped coordinates. In PIMD simulations, all beads corresponding to the same atom must share identical \texttt{image} flags during spring force calculations and normal mode transformations. However, LAMMPS updates the \texttt{image} flags independently for each bead during neighbor-list rebuilds via the \texttt{domain->pbc()} function. As a result, beads of the same atom may carry inconsistent \texttt{image} flags.

To resolve this issue, at the beginning of each PIMD step we use the \texttt{domain->unmap(x, image)} function to unwrap the coordinates of all beads to positions corresponding to an \texttt{image} flag of zero, without modifying the \texttt{image} flags themselves. After integrating the EOMs and transforming the normal mode coordinates back to Cartesian coordinates prior to force evaluation, we apply the \texttt{domain->unmap\_inv(x, image)} function to wrap the atomic coordinates back to positions consistent with their original \texttt{image} flags. In this way, the normal mode transformations and coordinate evolution are performed correctly, with all beads sharing the same periodic \texttt{image}. The calculation of properties such as $K_{\mathrm{CV}}$ is performed using similarly unwrapped coordinates.

\subsection{MPI Parallelization}
We now briefly discuss the parallelization strategy of our implementation. LAMMPS employs a partitioning mechanism to enable multi-replica simulations, such as nudged elastic band (NEB), parallel tempering, and PIMD~\cite{thompson_lammps_2022}. As discussed in Subsection~\ref{example_liquid}, a two-level Message Passing Interface (MPI) communicator is initialized. The top-level communication layer, referred to as the ``universe'', consists of $N = n \times M$ processors and is partitioned into $n$ ``worlds'', each corresponding to an MPI communicator comprising $M$ processes. Each group of $M$ processors performs the potential energy calculation for a single bead, and the $n$ partitions correspond to the $n$ beads. Inter-bead communication is required because both the normal mode transformation and the calculation of properties such as $P_{\mathrm{CV}}$ involve all beads associated with a given atom. Since each PIMD step requires three normal mode transformations, the inter-bead communication must be performed at least three times per step.

LAMMPS assigns each atom a unique \texttt{tag} that remains unchanged throughout the MD simulation, and all beads corresponding to the same atom share the \texttt{tag} value. However, per-atom quantities, such as coordinates and forces, associated with the same atom on different beads may be stored in different orders. When only one process is used for each bead, namely $M = 1$, we first sort the atomic coordinates (or forces) according to the \texttt{tag} values and then gather the sorted coordinates of all beads for all atoms using \texttt{MPI\_Allgatherv}. After performing the normal mode transformation, the transformed coordinates are reordered to match the original ordering on each bead.

\begin{figure*}
    \centering
    \includegraphics[width=1.0\linewidth]{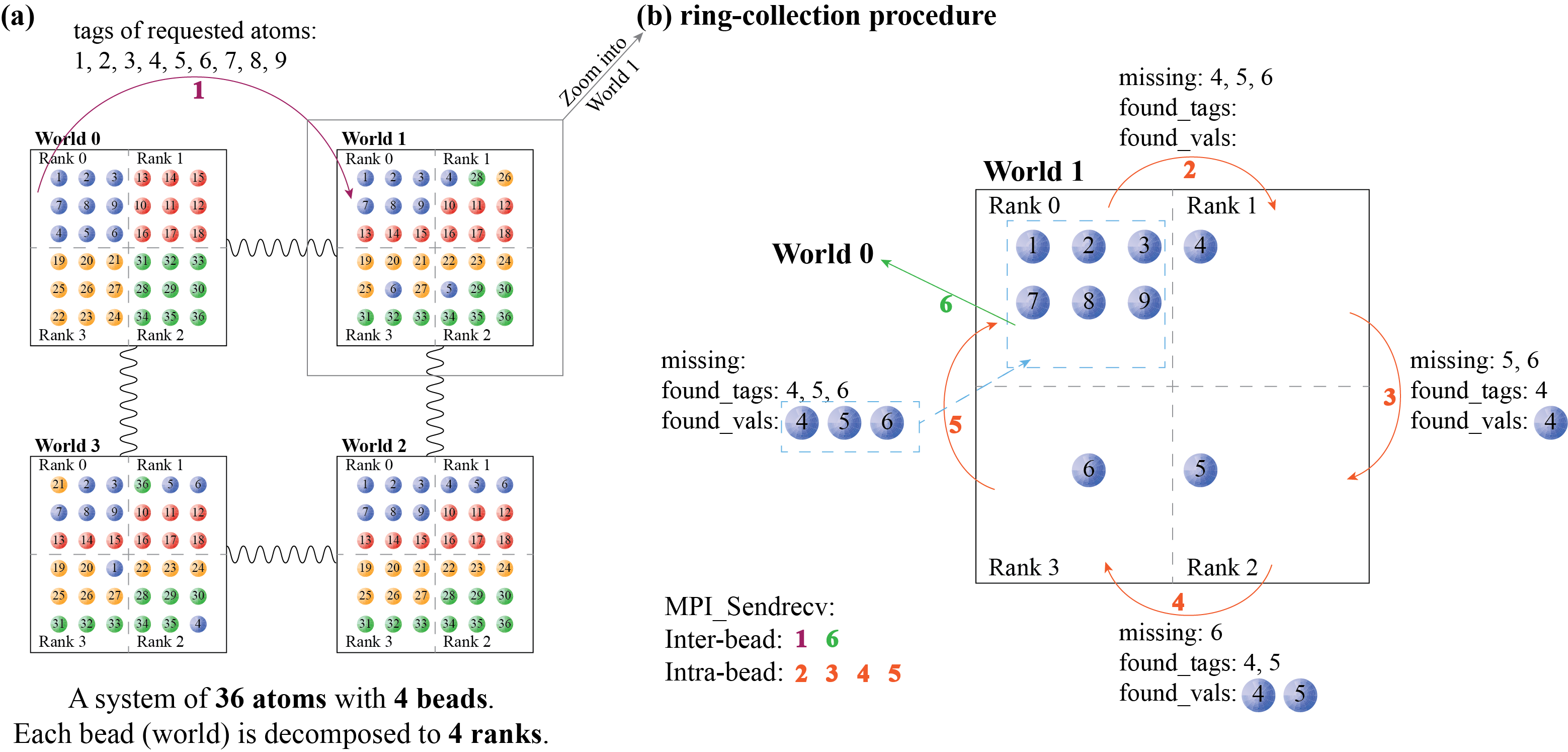}
    \caption{Inter-bead communication when multiple MPI processes are employed per bead. The colored balls denote per-atom quantities (either coordinates or forces) associated with the atom identified by the labeled \texttt{tag}. The schematic illustrates a system of 36 atoms with 4 beads using 4 MPI processes per bead. Each ``world'' denotes an MPI communicator composed of 4 ranks. Colored arrows denote \texttt{MPI\_Sendrecv} operations and the colored numbers indicate the order in which these operations are performed. Panel (b) shows a zoomed-in view of World 1 in panel (a) and illustrates the ``ring-collection'' procedure, in which Rank 0 of World 1 collects the atoms requested by Rank 0 of World 0. Atoms requested by other ranks are handled in the same manner and are omitted for clarity.}
    \label{fig:inter-beadcomm}
\end{figure*}
We now illustrate the parallelization strategy when multiple MPI processes are employed per bead, i.e., $M>1$. In this case, the atoms within each bead are distributed across multiple spatial domains. Because the domain decomposition is not guaranteed to be identical across different beads, a given atom may be assigned to different processor ranks in different beads. For example, in the case of 175,626 H$_2$O molecules with 32 beads, approximately 5\% of the atoms are not assigned to processors with the same rank in other beads. Consequently, a requesting rank cannot obtain the coordinates of these missing atoms from another bead by communicating solely with the rank of the same index in that bead. Instead, the target rank in the other bead must first collect all atoms requested by the requesting rank.

FIG.~\ref{fig:inter-beadcomm} illustrates this collection and communication procedure using a system of 36 atoms and 4 beads, showing how Rank~0 of World~0 retrieves the coordinates of its local atoms (the blue atoms with \texttt{tag}s 1-9) from World~1. First, as shown in FIG.~\ref{fig:inter-beadcomm} (a), Rank~0 of World~0 sends the tags of these atoms (1-9) to Rank~0 of World~1 (labeled by the purple arrow 1). In World~1, six of the nine atoms are local to Rank~0, while atoms 4, 5, and 6 reside on Ranks~1, 2, and 3, respectively. Afterwards, Rank~0 of World~1 employs an intra-bead ring-communication procedure to collect the coordinates of the remaining three atoms from the other ranks.

As shown in FIG.~\ref{fig:inter-beadcomm} (b), in each step of this ``ring-collection'' procedure, a rank searches for the missing atoms among its local atoms, copies the tags and coordinates of any atoms found, and removes the corresponding tags from the list of missing atoms. It then sends the updated list of missing tags, together with the tags and coordinates of the atoms found in this step, to the next rank in the ring (indicated by the orange arrows 2-5). After the search reaches Rank~3, all missing atoms have been found and their coordinates are sent back to Rank~0. In this way, Rank~0 of World~1 obtains the coordinates of all atoms requested by Rank~0 of World~0 and sends them back to World~0. Rank~0 of World~0 also collects the coordinates of these atoms from Worlds~2 and~3 using the same procedure. As a result, Rank~0 of World~0 obtains the coordinates of its local atoms in all beads and subsequently performs the normal mode transformation. The same procedure is applied to all other ranks. This collection-communication scheme minimizes the number of atoms communicated within each bead and exchanges only the necessary atomic data between beads, resulting in a relatively small amount of MPI communication and thereby ensuring efficient parallel performance.

\subsection{Bosonic PIMD}
To implement \texttt{fix pimd/langevin/bosonic}, a child class \texttt{FixPIMDBLangevin} is derived from the base class \texttt{FixPIMDLangevin}. The \texttt{method} parameter is set to \texttt{pimd}, which evolves the system in the Cartesian coordinates. The implementation uses the quadratic scaling algorithm~\cite{feldman_quadratic_2023}. In this approach, the spring potential is reformulated through a recurrence relation that includes contributions from rings combining several particles together. The forces are evaluated through expressions based on the probabilities that different particles are connected under the bosonic spring potential. The implementation follows the strategy of modifying the potential of beads $k=0$ and $k=n-1$, while the rest experience the same potential as in distinguishable particles (see the supporting information of Ref.~\citenum{higer_periodic_2025}).

\section{Performance}
In this section, we report the performance of \texttt{fix pimd/langevin}. Benchmark calculations are carried out for a liquid water system using the DP model described in Subsection~\ref{example_liquid}. All simulations are performed in the $NVT$ ensemble with a timestep of 0.5~fs. The DP models are evaluated using DeePMD-kit v2.2.9~\cite{zeng_deepmd-kit_2023}. We employ the 2 August 2023 stable release of LAMMPS in Subsection~\ref{bench_small} and the code developed in pull request 4857 in Subsection~\ref{bench_strong_weak}. All calculations are conducted on NERSC’s Perlmutter supercomputer, where each node is equipped with four NVIDIA A100 GPUs. To enable the inference of DP models for two beads simultaneously on a single GPU, NVIDIA’s Multi-Process Service (MPS) is used for the LAMMPS runs. For tests of i-PI, we use i-PI v3.1.5 connected with DeePMD-kit using its \texttt{dp\_ipi} interface. The performance of i-PI is evaluated across two protocols: UNIX-domain sockets for intra-node communication and TCP/IP sockets for inter-node scalability.
%Weile commented out.
%We benchmark the performance of i-PI using its two available communication mechanisms: the UNIX-domain socket, which provides local inter-process communication restricted to a single node, and the TCP/IP socket, an Internet-Protocol-based communication scheme that supports multi-node execution.

In Subsection~\ref{bench_small}, we first report benchmark results for two system sizes relevant to practical use cases, containing 512 and 4096 H$_2$O molecules, and compare the performance of \texttt{fix pimd/langevin} with that of i-PI. Then we discuss the strong and weak scaling behaviors of \texttt{fix pimd/langevin} by extending the benchmarks to extremely large systems and comparing with MD for classical atomic nuclei in Subsection~\ref{bench_strong_weak}. The computational efficiency is reported in terms of the execution time per MD step, where a wall-clock time of 1~s per step corresponds to a trajectory length of 0.043~ns/day.

\subsection{Performance at System Sizes for Practical Simulations}~\label{bench_small}
\begin{figure}
    \centering
    \includegraphics[width=\columnwidth]{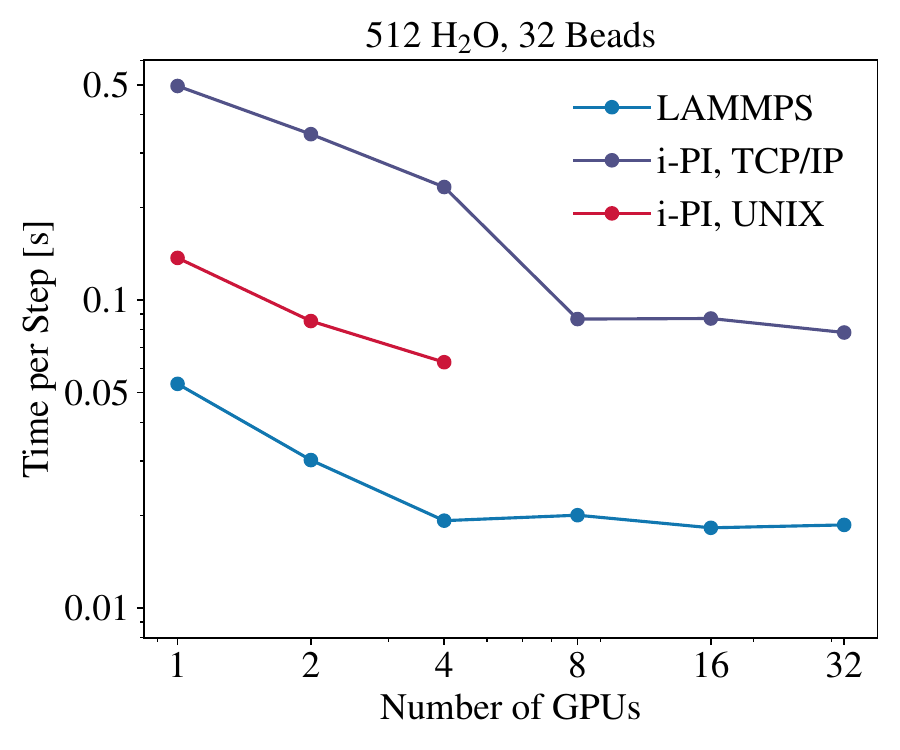}
    \caption{Benchmark of the performance of our LAMMPS implementation and i-PI for 512 H$_2$O molecules with 32 beads. Each node has four GPUs. UNIX communication is restricted to a single node, while TCP/IP supports multi-node execution.}
    \label{fig:512H2O}
\end{figure}

\begin{figure}
    \centering
    \includegraphics[width=\columnwidth]{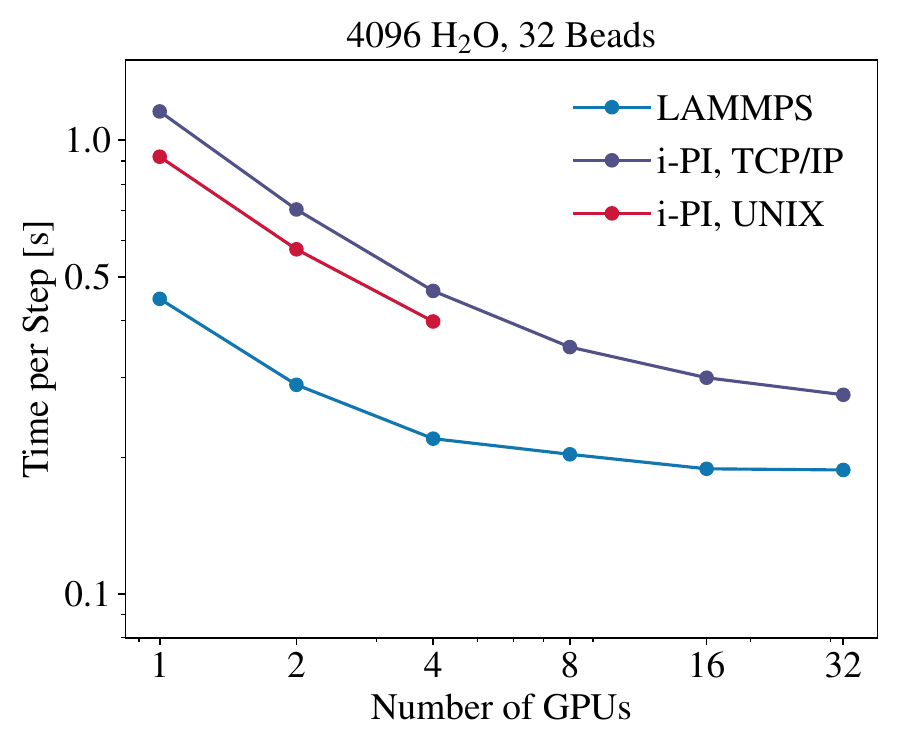}
    \caption{Benchmark of the performance of our LAMMPS implementation and i-PI for 4096 H$_2$O molecules with 32 beads. Each node has four GPUs. UNIX communication is restricted to a single node, while TCP/IP supports multi-node execution.}
    \label{fig:4096H2O}
\end{figure}

\begin{figure}
    \centering
    \includegraphics[width=\columnwidth]{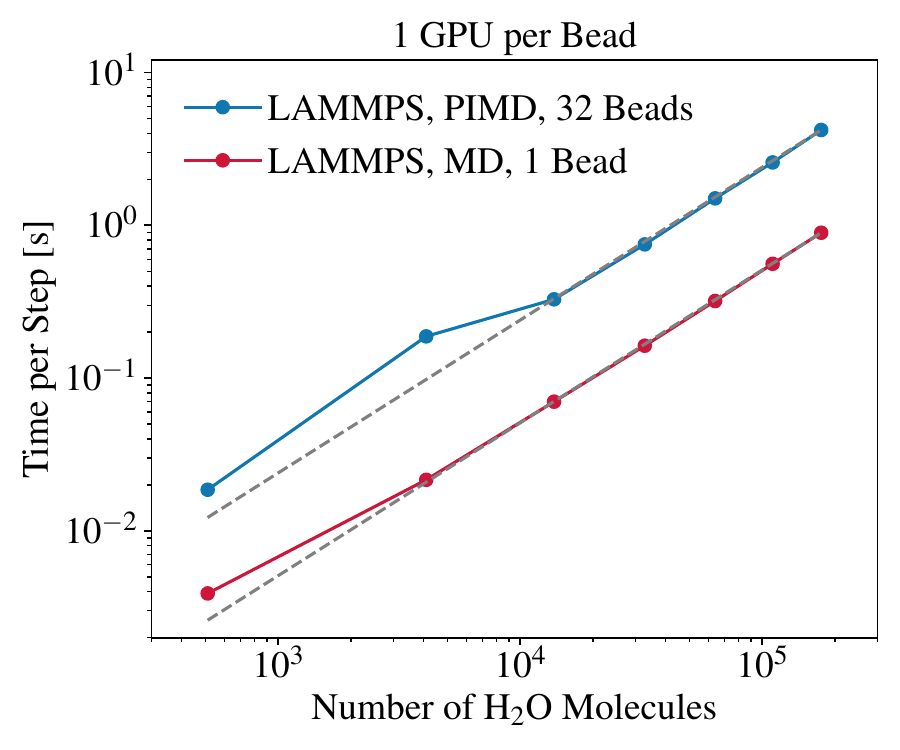}
    \caption{Scaling of our PIMD implementation with the number of atoms for a liquid-water system with 32 beads, compared with classical MD, which effectively corresponds to one bead. The gray lines indicate ideal scaling.}
    \label{fig:1gpuperbead}
\end{figure}

We first report benchmark results for a system of 512 H$_2$O molecules (FIG.~\ref{fig:512H2O}). The performance of our LAMMPS implementation is compared against i-PI using various communication protocols. The UNIX socket mechanism of i-PI is restricted to intra-node communication. Consequently, these tests were performed on a single node equipped with four GPUs. Our LAMMPS implementation achieves optimal efficiency when using all four GPUs on a single node,  reaching a speed of 2.25 ns/day (0.019 s/step). This is a 3.3-fold speedup over i-PI using UNIX sockets (0.69 ns/day) and a 12-fold speedup over i-PI via TCP/IP (0.18 ns/day). For this system size, our LAMMPS implementation reaches performance saturation at the single-node level; scaling beyond four GPUs causes inter-node communication overhead to outweigh the computational gains from additional hardware. In contrast, i-PI with TCP/IP achieves its peak performance of 0.50 ns/day (0.087 s/step) using two nodes (eight GPUs), reflecting the higher communication latency inherent in its client-server architecture.
%the best balance between the number of GPUs and performance when all four GPUs on a single node are used, yielding 0.019 s per step (2.25 ns/day). This speed is 3.3 times that of i-PI with UNIX sockets (0.69 ns/day) and 12 times that of i-PI with TCP/IP sockets (0.18 ns/day). When more than one node is used for the LAMMPS code, the communication cost dominates over the acceleration provided by additional GPUs, resulting in no further performance improvement. For i-PI with TCP/IP communication, the optimal performance is achieved using two nodes with a total of eight GPUs, yielding 0.087 s per step (0.50 ns/day).    

%Then we show benchmark results for a system of 4096 H$_2$O molecules in FIG.~\ref{fig:4096H2O}. A much smaller difference between the UNIX and TCP/IP communication mechanisms is observed, as communication between the integrator and the force engine plays a smaller role for larger systems. Our LAMMPS implementation still exhibits a significant acceleration compared to i-PI. When all four GPUs on a single node are used, our LAMMPS implementation yields a speed of 0.22 s per step (0.20 ns/day), corresponding to 1.8 times the speed of i-PI with UNIX sockets (0.11 ns/day) and 2.1 times the speed of i-PI with TCP/IP sockets (0.093 ns/day). When 32 GPUs are used, both our LAMMPS implementation and i-PI with TCP/IP reach their maximum speed, yielding 0.23 and 0.16 ns/day, respectively.

Benchmark results for a larger system of 4096 H$_2$O molecules are presented in FIG.~\ref{fig:4096H2O}. In this regime, the performance discrepancy between UNIX and TCP/IP protocols diminishes significantly. This is attributed to the increased compute-to-communication ratio: as the computational cost of force evaluations scales with system size, the fixed latency of the driver-client interface becomes a less dominant factor in the total wall-clock time. Our LAMMPS implementation continues to demonstrate superior efficiency. On a single node with four GPUs, it achieves a throughput of 0.20 ns/day (0.22 s/step), outperforming i-PI with UNIX and TCP/IP sockets by factors of 1.8 and 2.1, respectively. As we scale to 32 GPUs, both our implementation and i-PI via TCP/IP reach their strong-scaling limits, yielding peak throughputs of 0.23 and 0.16 ns/day, respectively. The convergence of these results indicates that for sufficiently large systems, the primary performance bottleneck shifts from the communication protocol to the inherent parallel efficiency of the underlying potential model and the MPI collective overhead associated with orchestrating a large number of MPI tasks. 

The system sizes in FIGs.~\ref{fig:512H2O} and~\ref{fig:4096H2O}, namely 512 and 4096 H$_2$O molecules, are of practical interest in simulations but are too small to take advantage of a large number of GPUs. The scaling behavior of our implementation with one GPU per bead (i.e., 32 GPUs in total) as a function of the number of H$_2$O molecules for a liquid-water system with 32 beads is shown in FIG.~\ref{fig:1gpuperbead}. %For comparison, classical MD exhibits nearly linear scaling for sufficiently large systems, inherit to the linear scaling behavior of the DP model. PIMD shows a similar linear dependence on system size, in agreement with the linear scaling of the inter-bead communication with the number of atoms. 
For comparison, classical MD exhibits nearly linear scaling for sufficiently large systems, which is consistent with the arithmetic characteristics of the DP model. When the system is sufficiently large (more than about $10^4$ H$_2$O molecules), our PIMD implementation maintains this linear dependence on system size, confirming that the inter-bead communication scales proportionally with the total number of atoms. Compared to classical MD, the additional inter-bead communication in PIMD introduces an approximately 370\% increase in simulation time.

\subsection{Strong Scaling Behavior}\label{bench_strong_weak}

\begin{figure}
    \centering
    \includegraphics[width=\columnwidth]{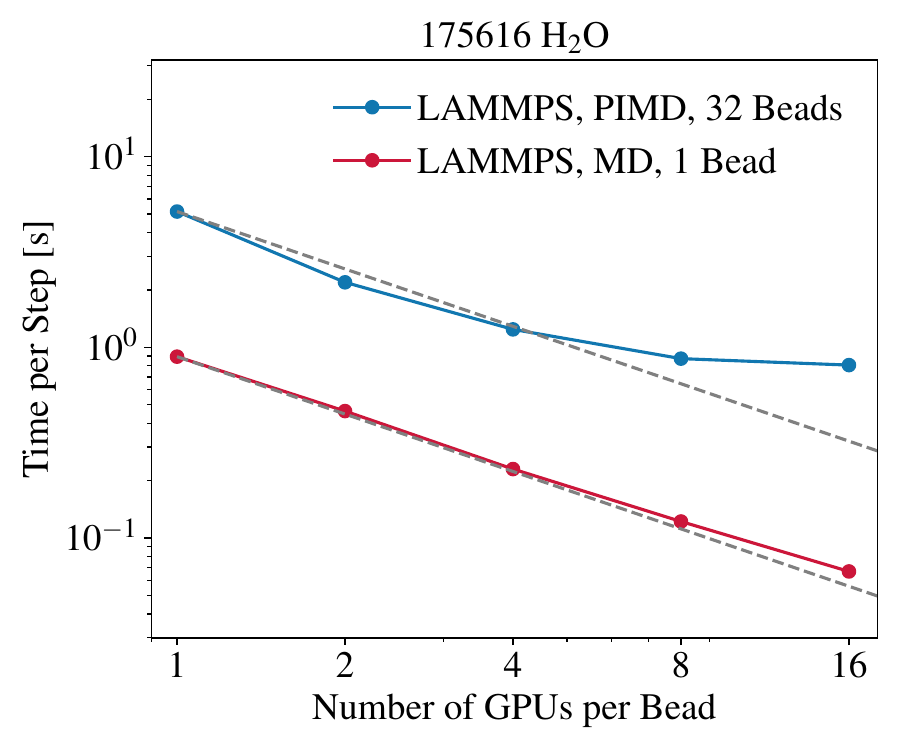}
    \caption{Strong scaling performance of our PIMD implementation for a system of 175,616 H$_2$O molecules with 32 beads, compared with classical MD, which effectively corresponds to 1 bead, for the same system. The gray lines indicate ideal scaling.}
    \label{fig:175616H2O}
\end{figure}

In FIG.~\ref{fig:175616H2O} we report the strong scaling behavior of our implementation using a system of 175,616 H$_2$O molecules with 32 beads, which is of the same order of magnitude as the maximum capacity of a single A100 GPU utilizing a compressed DP model, approximately $9.87\times10^{5}$ atoms~\cite{zeng_deepmd-kit_2023}. We evaluate our implementation against the strong scaling of the corresponding classical MD simulation, where the nuclei are represented by a single bead. The benchmark simulations start from one GPU per bead, and computations with multiple GPUs per bead are achieved by applying domain decomposition independently to each bead.

As shown in FIG.~\ref{fig:175616H2O}, the classical MD simulation exhibits near-ideal scaling up to 16 GPUs, indicating a low communication overhead within the domain-decomposition parallelization. The parallel efficiency of classical MD is fairly high, yielding 96.5\%, 97.1\%, 91.5\%, and 83.6\% when using 2, 4, 8, and 16 GPUs, respectively.

For PIMD with one GPU per bead, the performance decreases to 0.010~ns/day (4.2~s/step), compared to  0.048~ns/day (0.89~s/step) for classical MD. This 4.7-fold slowdown is attributed to the inter-bead communication overhead required for the three normal mode transformations performed at each PIMD step.

When multiple GPUs per bead are employed, our PIMD implementation exhibits near-ideal scaling up to 8 GPUs, with parallel efficiencies of 73.8\% and 39.9\% for 8 and 16 GPUs, respectively. This strong scaling behavior demonstrates that our implementation can substantially accelerate PIMD simulations of large systems by efficiently exploiting MPI parallelism.

\subsection{Weak Scaling Behavior}\label{bench_weak_weak}
\begin{figure}
    \centering
    \includegraphics[width=\columnwidth]{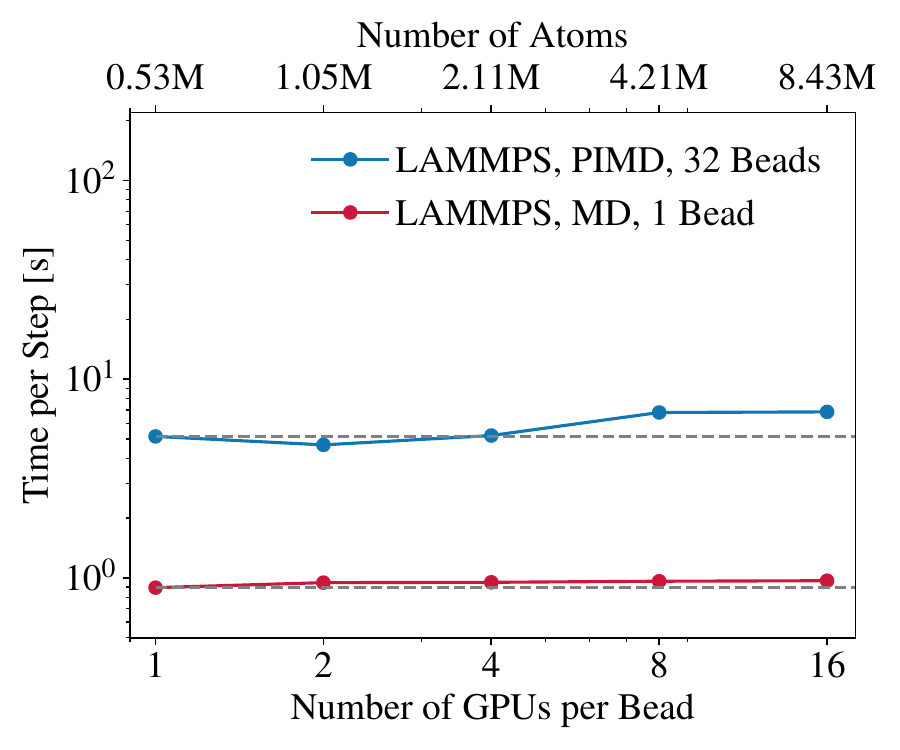}
    \caption{Weak scaling performance of our PIMD implementation for a liquid water system with 32 beads, compared with classical MD, which effectively corresponds to a single bead for the same system. Each GPU holds 175,616 H$_2$O molecules, corresponding to approximately 0.53 million atoms. As the number of GPUs per bead increases from 1 to 16, the total number of atoms increases from approximately 0.53M to 8.43M. The gray lines indicate ideal weak scaling.}
    \label{fig:weak}
\end{figure}

FIG.~\ref{fig:weak} shows the weak scaling behavior of our PIMD implementation. In these benchmarks, the workload on each GPU is kept constant at 175,616 H$_2$O molecules per GPU, culminating in a system of 8.43 million atoms for the largest PIMD simulation. In comparison, classical MD exhibits nearly ideal weak scaling up to 8.43 million atoms, reaching a parallel efficiency of 92.3\% for 16 GPUs, a result consistent with previous high-performance benchmark results reported in Ref.~\citenum{jia_pushing_2020}.

For PIMD, the weak scaling tests show parallel efficiencies of 89.9\%, 80.7\%, 61.8\%, and 61.3\% for 2, 4, 8, and 16 GPUs per bead, respectively. The increase in execution time with a larger number of GPUs per bead can be attributed to the overhead of intra-bead communication associated with the atom-collection procedure illustrated in Fig.~\ref{fig:inter-beadcomm}(b). Nevertheless, the observed weak scaling behavior demonstrates the strong utility of our implementation for large-scale PIMD simulations.

% weak scaling remains efficient up to two GPUs per bead, with the wall-clock time increasing by only 32\% when increasing from one to two GPUs per bead. However, as the number of GPUs and atoms are scaled further, the execution time per step grows steadily. This behavior can be attributed to the current implementation of inter-bead communication, in which all atoms of each bead are first collected and then exchanged among beads. These results highlight that while the computational cost per GPU remains fixed, the communication complexity associated with this all-to-all synchronization limits weak scaling. Transitioning to a more localized or asynchronous communication protocol may be necessary to further enhance the large-scale performance of our implementation. 
%These results indicate that further optimization of the inter-bead communication may improve the weak scaling performance of our implementation.

\section{Conclusions}
We have presented \texttt{fix pimd/langevin}, a highly efficient implementation of PIMD in LAMMPS, together with complete hands-on usage examples. The implementation provides commonly used PIMD features, including simulations in the $NVE$, $NVT$, and $NpT$ ensembles, as well as useful output properties such as $K_{\mathrm{CV}}$. The correctness of our implementation has been rigorously validated against i-PI. For system sizes relevant to practical simulations, our implementation achieves several-fold acceleration compared to i-PI. In addition, \texttt{fix pimd/langevin} exhibits good strong and weak scaling behaviors. Overall, this implementation provides a valuable tool for atomistic-scale modeling of nuclear quantum effects and constitutes a meaningful addition to the LAMMPS software ecosystem.

\section*{Acknowledgments}
The authors thank Chris Knight and Yuxing Peng from the Voth group at University of Chicago for contributing the first PIMD implementation in LAMMPS in October 2011. The authors also thank Linfeng Zhang, Xinzijian Liu, Han Wang, Yixiao Chen, Pinchen Xie, Bingjia Yang, Jinzhe Zeng, Han Bao, Zhuoqiang Guo, Jianxiong Li, and Mark Tuckerman for helpful discussions. Y.L., A.G., K.C., and R.C. acknowledge support from the Computational Chemical Sciences Center ``Chemistry in Solution and at Interfaces'' under Award No. DE-SC0019394 from the U.S. Department of Energy. Y.M.Y.F. was supported by Schmidt Science Fellows, in partnership with the Rhodes Trust. O.B. acknowledges support from the Clore Scholars Programme of the Clore Israel Foundation. B.H. acknowledges support from the Israel Science Foundation (grants No. 1037/22 and 1312/22). Financial support provided to A.K. by Sandia National Laboratories under POs~2149742 and 2407526 is gratefully acknowledged. Resources of the National Energy Research Scientific Computing Center (DoE Contract No. DE-AC02-05cH11231) were used. The authors also acknowledge use of Princeton University’s Research Computing resources.

\section*{AUTHOR DECLARATIONS}
\subsection*{Conflict of Interest}
The authors have no conflicts to disclose.
\subsection*{Author Contributions}
\textbf{Yifan Li}: Conceptualization (lead); Data Curation (lead); Software (lead); Visualization (lead); Writing - original draft preparation (lead); Writing - review \& editing (equal).
\textbf{Axel Gomez}: Conceptualization (supporting); Data Curation (supporting); Visualization (supporting); Writing - Original Draft Preparation (supporting); Writing - Review \& Editing (equal).
\textbf{Kehan Cai}: Data Curation (supporting); Writing - Original Draft Preparation (supporting); Writing - Review \& Editing (equal).
\textbf{Chunyi Zhang}: Data Curation (supporting); Writing - Original Draft Preparation (supporting); Writing - Review \& Editing (equal).
\textbf{Li Fu}: Software (supporting); Writing - Original Draft Preparation (supporting); Writing - Review \& Editing (equal).
\textbf{Weile Jia}: Conceptualization (supporting); Writing - Review \& Editing (equal).
\textbf{Yotam M. Y. Feldman}: Software (supporting); Writing - Original Draft Preparation (supporting); Writing - Review \& Editing (equal).
\textbf{Ofir Blumer}: Software (supporting); Writing - Review \& Editing (equal).
\textbf{Jacob Higer}: Software (supporting); Writing - Review \& Editing (equal).
\textbf{Barak Hirshberg}: Software (supporting); Writing - Review \& Editing (equal).
\textbf{Shenzhen Xu}: Supervision (equal); Writing - Review \& Editing (equal).
\textbf{Axel Kohlmeyer}: Software (supporting); Writing - Review \& Editing (equal).
\textbf{Roberto Car}: Supervision (equal); Writing - Review \& Editing (equal).

\section*{DATA AVAILABILITY}
The data that support the findings of this study are openly available in GitHub at https://github.com/lammps/lammps and Zenodo at https://doi.org/10.5281/zenodo.18422744.

\appendix
\section{Output Properties}\label{app:properties}
Here we provide a complete list of the mathematical expressions for the output properties. In the following, $N$ denotes the number of atoms, $T$ the temperature, and $n$ the number of beads. The frequency of the harmonic springs, $\omega_n$, is defined in Eq.~\eqref{omega_n}.

In the following, the superscript $(k)$ denotes the $k$th bead or normal mode, with $k=0,\ldots,n-1$, and the subscript $i$ denotes the $i$th atom, with $i=1,\ldots,N$. The vectors $\bm{r}_{i}^{(k)}$, $\bm{v}_{i}^{(k)}$, and $\bm{F}_{i}^{(k)}$ are the Cartesian coordinate, velocity, and force of the $k$th bead of the $i$th atom, respectively. The centroid of the beads associated with the $i$th atom is defined as $\bm{r}_{i}^{(c)} = \frac{1}{n} \sum_{k=0}^{n-1} \bm{r}_{i}^{(k)}$. The vectors $\tilde{\bm{r}}_{i}^{(k)}$, $\tilde{\bm{v}}_{i}^{(k)}$, and $\tilde{\bm{F}}_{i}^{(k)}$ are the normal mode coordinate, velocity, and force of the $k$th bead of the $i$th atom, respectively.

\begin{enumerate}[series=props]
\item Kinetic Energy
\begin{itemize}[leftmargin=0pt]
\item For a normal mode (when \texttt{method} is \texttt{nmpimd}):
\begin{equation}
K^{(k)}=\sum_{i=1}^{N}\frac{1}{2}m_i\tilde{\bm{v}}_{i}^{(k)2}.
\end{equation}
\item For a bead in the Cartesian coordinate (when \texttt{method} is \texttt{pimd}):
\begin{equation}
\label{eq:kinetic-energy-of-the-bead-property}
K^{(k)}=\sum_{i=1}^{N}\frac{1}{2}m_i\bm{v}_{i}^{(k)2}.
\end{equation}
\end{itemize}

\item Spring Elastic Energy 
\begin{itemize}[leftmargin=0pt]
\item For a normal mode (when \texttt{method} is \texttt{nmpimd}):
\begin{equation}
S^{(k)}=\sum_{i=1}^{N}\frac{1}{2}m_i\omega_n\lambda_k\tilde{\bm{r}}_{i}^{(k)2}
\end{equation}
where $\lambda_k$ is defined in Eq.~\eqref{lambda_k}.
\item For a bead in the Cartesian coordinate (when \texttt{method} is \texttt{pimd}):
\begin{equation}
\label{eq:spring-elastic-energy-of-the-bead-property}
S^{(k)}=\sum_{i=1}^{N}\frac{1}{4}m_i\omega_n\left[(\bm{r}_{i}^{(k)}-\bm{r}_{i}^{(k-1)})^2+(\bm{r}_{i}^{(k)}-\bm{r}_{i}^{(k+1)})^2\right]
\end{equation}
where $\bm{r}_{i}^{(n)}=\bm{r}_{i}^{(0)}$.
\end{itemize}

\item Potential Energy of the Bead
\begin{equation}
\label{eq:potential-energy-of-the-bead-property}
U^{(k)}(\bm{r}_{1}^{(k)},\dots,\bm{r}_{N}^{(k)}).
\end{equation}
\end{enumerate}

Outputs 1, 2, and 3 differ in the log files corresponding to each bead, while all subsequent outputs are identical in the log files of all beads.

\begin{enumerate}[resume=props]
\item Total Energy of All Beads
\begin{equation}\label{tot_energy}
H=\sum_{k=0}^{n-1}\left(K^{(k)}+S^{(k)}+U^{(k)}\right).
\end{equation}

\item Primitive Kinetic Energy Estimator
\begin{equation}\label{eq_kpr}
K_{\mathrm{PR}}=\frac{3nNk_{\mathrm{B}}T}{2}-\sum_{k=0}^{n-1}\sum_{i=1}^{N}\frac{1}{2}m_i\omega_n(\bm{r}_{i}^{(k)}-\bm{r}_{i}^{(k+1)})^2.
\end{equation}

\item Virial Kinetic Energy Estimator
\begin{equation}\label{eq_vir}
K_{\mathrm{VR}}=-\frac{1}{2n}\sum_{k=0}^{n-1}\sum_{i=1}^{N}\bm{r}_{i}^{(k)}\cdot \bm{F}_{i}^{(k)}.
\end{equation}
$K_{\mathrm{VR}}$ should not be used for simulations under PBCs, whereas $K_{\mathrm{PR}}$ and $K_{\mathrm{CV}}$ are the appropriate choice. $K_{\mathrm{CV}}$ is recommended because it has smaller fluctuations than $K_{\mathrm{PR}}$~\cite{tuckerman_statistical_2023}.

\item Centroid-Virial Kinetic Energy Estimator
\begin{equation}\label{eq_kcv}
K_{\mathrm{CV}}=\frac{3Nk_{\mathrm{B}}T}{2}-\frac{1}{2n}\sum_{k=0}^{n-1}\sum_{i=1}^{N}\left(\bm{r}_{i}^{(k)}-\bm{r}_{i}^{(c)}\right)\cdot \bm{F}_{i}^{(k)}.
\end{equation}

\item Primitive Pressure Estimator
\begin{equation}\label{eq_ppr}
\begin{aligned}
P_{\mathrm{PR}}=&\frac{1}{3V}\left(3nNk_{\mathrm{B}}T-\sum_{k=0}^{n-1}\sum_{i=1}^{N}m_i\omega_n(\bm{r}_{i}^{(k)}-\bm{r}_{i}^{(k+1)})^2
\right).
\end{aligned}
\end{equation}

Given the cell tensor $\left\{h_{\alpha\beta}\right\}$ with $h_{\alpha\beta}$ being the $\beta$-th component of the $\alpha$-th cell vector, $\bm{\Xi}_{\alpha\beta}^{(k)}=-\sum_{\gamma}\frac{\partial U^{(k)}}{\partial h_{\alpha\gamma}}h_{\gamma\beta}$ represents the virial of the $k$-th bead.

\item Pressure of the Extended Classical System
\begin{equation}
P_{\mathrm{MD}}=\frac{1}{3V}\sum_{k=0}^{n-1}\left(\sum_{i=1}^{N}m_i\tilde{\bm{v}}_{i}^{(k)2}+\mathrm{Tr}\left[\bm{\Xi}^{(k)}\right]\right).
\end{equation}

\item Centroid-Virial Pressure Estimator
\begin{equation}\label{eq_pcv}
\begin{aligned}
P_{\mathrm{CV}}=&\frac{1}{3nV}\left(\sum_{i=1}^{N}m_i\tilde{\bm{v}}_{i}^{(0)2}-\sum_{k=0}^{n-1}\sum_{i=1}^{N}\left(\bm{r}_{i}^{(k)}-\bm{r}_{i}^{(c)}\right)\cdot \bm{F}_{i}^{(k)}\right.\\
&\left.+\sum_{k=0}^{n-1}\mathrm{Tr}\left[\bm{\Xi}^{(k)}\right]\right).
\end{aligned}
\end{equation}

\item Barostat Velocity $v_{\mathrm{W}}$ (isotropic barostat) or $v_{\mathrm{W}\alpha}, \alpha=x, y, z$ (anisotropic barostat), as described in Appendix~\ref{BZP_barostat}.

\item Barostat Kinetic Energy
\begin{itemize}[leftmargin=0pt]
\item Isotropic barostat
\begin{equation}\label{baro_ke}
K_{\mathrm{W}}=\frac{1}{2}Wv_{\mathrm{W}}^2\quad 
\end{equation}
where the barostat mass is $W=3nNk_{\mathrm{B}}T\tau_P^2$ with the damping time $\tau_P$ given by parameter \texttt{taup}.
\item Anisotropic barostat
\begin{equation}
K_{\mathrm{W}}=\sum_{\alpha=x,y,z}\frac{1}{2}Wv_{\mathrm{W}\alpha}^2\quad
\end{equation}
where $W=nNk_{\mathrm{B}}T\tau_P^2$.
\end{itemize}

\item Barostat Potential Energy
\begin{equation}\label{baro_pe}
   U_{\mathrm{W}}=nP_{\mathrm{ext}}V
\end{equation}
where $P_{\mathrm{ext}}$ is the target pressure of the barostat.
\item Barostat Cell Jacobian
\begin{equation}\label{baro_jaco}
    J_{\mathrm{W}}=-n k_{\mathrm{B}}T\ln(V).
\end{equation}
\item Enthalpy of the Extended System
\begin{equation}\label{total_enthalpy}
\mathcal{H}=H+K_{\mathrm{W}}+U_{\mathrm{W}}+J_{\mathrm{W}}
\end{equation}
where $H$, $K_{\mathrm{W}}$, $U_{\mathrm{W}}$, and $J_{\mathrm{W}}$ are defined in Eqs.~\eqref{tot_energy}, ~\eqref{baro_ke}, ~\eqref{baro_pe}, ~\eqref{baro_jaco}, respectively.
\end{enumerate}

The properties output by \texttt{fix pimd/langevin/bosonic} are as follows:
\begin{enumerate}[series=props-bosons]
\item Kinetic Energy of the Bead: the same as Eq.~\eqref{eq:kinetic-energy-of-the-bead-property}.
% \begin{equation}
% K^{(k)}=\sum_{i=1}^{N}\frac{1}{2}m_i\bm{v}_{i}^{(k)2}.
% \end{equation}

\item Spring Elastic Energy of the Bead: 
\begin{itemize}[leftmargin=0pt]
\item For beads $k=1,\ldots,n-1$: the same as Eq.~\eqref{eq:spring-elastic-energy-of-the-bead-property}.
\item For bead $k=0$: the potential that involves a sum over ring polymer configurations with exchange, as described in Eq.~(3) of Ref~\citenum{feldman_quadratic_2023}.
\end{itemize}

\item Potential Energy of the Bead: the same as Eq.~\eqref{eq:potential-energy-of-the-bead-property}.
% \begin{equation}
% U^{(k)}(\bm{r}_{1}^{(k)},\dots,\bm{r}_{N}^{(k)}).
% \end{equation}
\end{enumerate}

Outputs 1, 2, and 3 differ in the log files corresponding to each bead, while all subsequent outputs are identical in the log files of all beads.

\begin{enumerate}[resume=props-bosons]
\item Total Energy of All Beads: the same as Eq.~\eqref{tot_energy}.
% \begin{equation}\label{tot_energy}
% H=\sum_{k=0}^{n-1}\left(K^{(k)}+S^{(k)}+U^{(k)}\right).
% \end{equation}

\item Primitive Kinetic Energy Estimator: the bosonic estimator as described in Eq.~(4) of the Supporting Information of Ref.~\citenum{hirshberg_path_2019}.
% \begin{equation}\label{eq_kpr}
% K_{\mathrm{PR}}=\frac{3nNk_{\mathrm{B}}T}{2}-\sum_{k=0}^{n-1}\sum_{i=1}^{N}\frac{1}{2}m_i\omega_n(\bm{r}_{i}^{(k)}-\bm{r}_{i}^{(k+1)})^2.
% \end{equation}

\item Virial Kinetic Energy Estimator: the same as Eq.~\eqref{eq_vir}, which is also valid in bosons~\cite{hirshberg_path_2020}.
% \begin{equation}\label{eq_vir}
% K_{\mathrm{VR}}=-\frac{1}{2n}\sum_{k=0}^{n-1}\sum_{i=1}^{N}\bm{r}_{i}^{(k)}\cdot \bm{F}_{i}^{(k)}.
% \end{equation}
$K_{\mathrm{VR}}$ should not be used for simulations under PBCs, whereas $K_{\mathrm{PR}}$ is the appropriate choice.

\end{enumerate}

\section{Normal Mode Transformation}\label{app:normalmode}
The Cartesian coordinate is transformed into the normal mode coordinate via the matrix $M$ defined as~\cite{martyna_molecular_1999}
\begin{equation}\label{Mij}
\begin{aligned}
M_{ij}=&
\begin{cases}
\dfrac{1}{\sqrt{n}},
& i=0,\\[8pt]
\dfrac{1}{\sqrt{n}}(-1)^j,
& i=\dfrac{n}{2}\ \text{and}\ n\ \text{even},\\[10pt]
\sqrt{\dfrac{2}{n}}
\cos\!\left(\dfrac{2\pi i j}{n}\right),
& 1 \le i \le \left\lfloor \dfrac{n}{2} \right\rfloor,\\[10pt]
\sqrt{\dfrac{2}{n}}
\sin\!\left(\dfrac{2\pi i j}{n}\right),
& \left\lfloor \dfrac{n}{2} \right\rfloor + 1 \le i \le n-1,
\end{cases}
% \\[0pt]
\\
&j=0, \dots, n-1,
\end{aligned}
\end{equation}
where $i$ indicates the normal modes and $j$ indicates the beads, and $n$ is the total number of beads.

The eigenvalues of the matrix $M$ are
\begin{equation}\label{lambda_k}
\lambda_k=4\sin^2\left(\frac{k\pi}{n}\right), k=0,\dots,n-1.
\end{equation}

Since the matrix $M$ defined in Eq.~(A1) is orthogonal, the inverse transformation is simply given by
\begin{equation}
M^{-1}_{ij} = M_{ji}.
\end{equation}

\section{de Boer Quantumness Parameter}\label{deboer}
For a system with the mass $m$, the length scale $\sigma$ and the energy scale $\epsilon$, the de Boer Quantumness parameter~\cite{de_boer_chapter_1957} is defined as
\begin{equation}
\Lambda^{*} = h \big/ \left( \sigma \sqrt{m \epsilon} \right)
\end{equation}
where $h$ is Planck’s constant. For example, the Neon system can be described using $m = 20.1797$ g/mol, $\epsilon = 3.0747 \times 10^{-3}$ eV, and $\sigma = 2.7616$~\AA. Then the quantumness is
\begin{equation}
\begin{aligned}
\Lambda^{*}=&4.135667403 \times 10^{-3}\ \mathrm{eV\cdot ps}/\Big(2.7616~\text{\AA}\\
&\times\sqrt{20.1797~\text{g/mol}\times3.0747 \times 10^{-3}~\mathrm{eV}}\\
&\times \sqrt{1.0364269 \times 10^{-4}~\mathrm{eV^{-1}\cdot mol\cdot g^{-1}\cdot \text{\AA}^{-2}\cdot ps^{2}}}\Big)\\
=&0.600.
\end{aligned}
\end{equation}
Thus, for a fully quantum simulation of Neon using \texttt{lj} units with the parameters stated above, \texttt{sp} should be set to 0.600.

\section{Integrator of the PILE\_L Thermostat}\label{PILE_L_thermostat}
In the PILE\_L thermostat for the normal mode PIMD~\cite{ceriotti_efficient_2010}, the damping time for the centroid mode $\tau_0$ can be set by the parameter \texttt{tau} and has a default value of 1.0 in the user-defined time unit. The default damping time $\tau_k$ for each non-centroid mode $k$ is
\begin{equation}
\tau_k=\frac{1}{2\omega_k}, k=1, ..., n-1
\end{equation}
where $\omega_k=\sqrt{\lambda_k}\omega_n$ with $\lambda_k$ and $\omega_n$ defined in Eqs.~\eqref{lambda_k} and~\eqref{omega_n}, respectively. The parameter \texttt{scale} is used to scale $\tau_k$ with $k>0$. 

The normal mode velocity is updated via
\begin{equation}\label{pile_l_eom}
\tilde{v}_{i\alpha}^{(k)}(t+\Delta t)=c_1^{(k)}\tilde{v}_{i\alpha}^{(k)}(t)+c_2^{(k)}\sqrt{\frac{n}{m_i\beta}}\eta_{i\alpha}^{(k)},\quad\alpha=x,y,z,
\end{equation}
where $c_1^{(k)}=e^{-\Delta t/\tau_k}$, $c_2^{(k)}=\sqrt{1-c_1^{(k)2}}$, and $\eta_{i\alpha}^{(k)}\sim\mathcal{N}(0, 1)$ is an independent random number following the standard Gaussian distribution.

For PIMD in the Cartesian coordinate, the default damping time is $\tau_k=\frac{1}{2\omega_n}, k=0, ..., n-1$. The Cartesian velocity is updated via
\begin{equation}\label{pile_l_eom}
v_{i\alpha}^{(k)}(t+\Delta t)=c_1v_{i\alpha}^{(k)}(t)+c_2\sqrt{\frac{n}{m_i\beta}}\eta_{i\alpha}^{(k)},\quad\alpha=x,y,z,
\end{equation}
where $c_1=e^{-\Delta t/\tau_0}$ and $c_2=\sqrt{1-c_1^{2}}$.

\section{Integrator of the BZP Barostat}\label{BZP_barostat}
For an isotropic barostat with a target pressure $P_{\mathrm{ext}}$ and mass $W$, the velocity $v_{\mathrm{W}}$, which is equal to $v_{\mathrm{W}\alpha}, \alpha=x, y, z$, is updated according to~\cite{bussi_isothermal-isobaric_2009}
\begin{equation}
\begin{aligned}
v_{\mathrm{W}}(t+\Delta t)=&v_{\mathrm{W}}(t)+\frac{3\Delta t}{W}\left[nV(P_{\mathrm{CV}}-P_{\mathrm{ext}})+\frac{1}{nk_{\mathrm{B}}T}\right]\\
&+\frac{\Delta t^2}{W}\sum_{i=1}^{N}\tilde{\bm{F}}_{i}^{(0)}\cdot\tilde{\bm{v}}_{i}^{(0)}+\frac{\Delta t^3}{3W}\sum_{i=1}^{N}\frac{||\tilde{\bm{F}}_{i}^{(0)}||^2}{m_i}.
\end{aligned}
\end{equation}

For an anisotropic barostat with a target pressure $P_{\mathrm{ext}}$ on dimension $\alpha$, the velocity $v_{\mathrm{W}\alpha}$ is updated as
\begin{equation}
\begin{aligned}
v_{\mathrm{W}\alpha}(t+\Delta t)=&v_{\mathrm{W}\alpha}(t)+\frac{\Delta t}{W}\left[nV(P_{\mathrm{CV}}^{\alpha\alpha}-P_{\mathrm{ext}})+\frac{1}{nk_{\mathrm{B}}T}\right]\\
&+\frac{\Delta t^2}{W}\sum_{i=1}^{N}\tilde{F}_{i\alpha}^{(0)}\tilde{v}_{i\alpha}^{(0)}+\frac{\Delta t^3}{3W}\sum_{i=1}^{N}\frac{\tilde{F}_{i\alpha}^{(0)2}}{m_i}
\end{aligned}
\end{equation}
where $P_{\mathrm{CV}}^{\alpha\beta}$ is the $\alpha\beta$ component of the centroid-virial stress tensor, defined as
\begin{equation}
\begin{aligned}
P_{\mathrm{CV}}^{\alpha\beta}=&\frac{1}{nV}\left(\sum_{i=1}^{N}m_i\tilde{v}_{i\alpha}^{(0)}\tilde{v}_{i\beta}^{(0)}-\sum_{k=0}^{n-1}\sum_{i=1}^{N}\left(r_{i\alpha}^{(k)}-r_{i\alpha}^{(c)}\right)\cdot F_{i\beta}^{(k)}\right.\\
&\left.+\sum_{k=0}^{n-1}\bm{\Xi}^{(k)}_{\alpha\beta}\right).
\end{aligned}
\end{equation}

Given the barostat velocity $v_{\mathrm{W}\alpha}$, the normal mode positions and velocities of the atoms are updated as
\begin{equation}
\tilde{r}_{i\alpha}(t+\Delta t)
=
e^{\Delta t v_{\mathrm{W}\alpha}}\,\tilde{r}_{i\alpha}(t)
+
\frac{
e^{\Delta t v_{\mathrm{W}\alpha}} - e^{-\Delta t v_{\mathrm{W}\alpha}}
}{2\,v_{\mathrm{W}\alpha}}
\,\tilde{v}_{i\alpha}(t),
\end{equation}
and
\begin{equation}
\tilde{v}_{i\alpha}(t+\Delta t)
=
e^{-\Delta t v_{\mathrm{W}\alpha}}\,\tilde{v}_{i\alpha}(t).
\end{equation}
The cell length $L_{\alpha}$ on dimension $\alpha$ is updated as
\begin{equation}
L_{\alpha}(t+\Delta t)=e^{\Delta t v_{\mathrm{W}\alpha}}L_{\alpha}(t).
\end{equation}

To sample the $NpT$ ensemble, the barostat velocity can be thermostatted as
\begin{equation}
v_{\mathrm{W}\alpha}(t+\Delta t)
=
c_1\, v_{\mathrm{W}\alpha}(t)
+
c_2\sqrt{\frac{nk_{\mathrm{B}}T}{m_i}}\eta_\alpha
\end{equation}
where $c_1=e^{-\Delta t/\tau_0}$, $c_2=\sqrt{1-c_1^{2}}$, and $\eta_{\alpha}\sim\mathcal{N}(0, 1)$ is an independent random number following the standard Gaussian distribution.
\bibliography{clean}% Produces the bibliography via BibTeX.

% \printbibliography

\end{document}